# Systematic study of α decay using different versions of proximity formalism

O. N. Ghodsi and A. Daei-Ataollah[*]

*Department of Physics, Faculty of Sciences, University of Mazandaran, P. O. Box 47415-416, Babolsar, Iran*

Finding the best model to describe the α-decay process is an old and ongoing challenge in nuclear physics. The present work systematically studied α-decay half-lives for the favored ground-state-to-ground-state transitions of 344 isotopes of nuclei with $52 \leq Z \leq 107$ using 28 versions of the proximity potential model in the framework of the WKB approximation. The present study introduces the best proximity versions with the fewest deviations with respect to experimental values. The models for Prox. 77-set 4, Prox. 77-set 5, and Dutt 2011 with the root-mean-square deviations (RMSDs) of <1 were found to predict α-decay half-lives better than the other models. Comparison with fusion studies shows that Dutt 2011 is an appropriate model both for α-decay studies and for prediction of the barrier characteristic in heavy-ion fusion reactions. The calculation of α-decay half-lives were repeated for even-even, even-odd, odd-even, and odd-odd nuclei. This detailed comparative study reveals that for these versions the half-lives of the even-even nuclei with RMSDs of <0.6 show less deviation than the even-odd, odd-even, and odd-odd nuclei.

## I. INTRODUCTION

Phenomenological proximity potential was first proposed by Blocki et al. [1] for heavy-ion reactions. This well-known applicable model with simple and accurate formalism has the advantage of adjustable parameters. This model requires the shape and the geometry of the participant nuclei and the universal function related to surface separation distance for formulation. Modifications to parameters in the original proximity potential have been made over time to generalize this model to fusion reactions and to overcome its shortcomings; these include the surface energy coefficient, surface thickness parameter, nuclei radius, and the universal function. Various versions of the original model now exist [2-8], although in some cases the modifications are minor [9-24].

It is known that α decay proceeds in the opposite direction of fusion between an α particle and a daughter nucleus; thus, the same interaction potential can be used to describe both processes. Of the various versions of the proximity model formulated for fusion studies, a few have been applied to α-decay studies [25-31]. Reviews of the performance of proximity models for predicting fusion cross sections exist [18,20,32], but thus far there has been no similar study for α-decay half-lives. The α-decay studies are usually limited to a few proximity model versions or to a restricted range of α emitters. Finding the best model/models for prediction of both fusion and decay properties would be beneficial and time-saving.

The present study extends the calculations to a wide range of proximity versions and more α emitters. The main objective is to investigate the best proximity model for proper prediction of α-decay half-lives. A detailed, comparative, systematic study of 28 different versions of the proximity model has been carried out to achieve this goal. The half-lives of α decay from ground-state-to-ground-state transitions of 344 nuclei with $52 \leq Z \leq 107$ have been calculated in the framework of the Wentzel-Kramers-Brillouin (WKB) [33] approximation. The best versions of the proximity potential for prediction of α-decay half-lives were chosen through comparison with experimental results and with the results of other approaches. Comparison with fusion studies has been used to find the most appropriate model for both fusion and for α-decay studies.

The formalism employed to calculate the α-decay half-lives and a detailed description of the total interaction potential between an α particle and a daughter nucleus is given in Sec. II with a focus on the nuclear proximity potential and its versions. The results and a discussion are presented in Sec. III. That section also makes comparisons with other models. The conclusions and direction of future study are discussed in Sec. IV.

* aysan.daei @gmail.com

## II. THEORETICAL FRAMEWORK

### A. Description of α-decay half-life formalism

The half-life of the parent nucleus opposed to fission to an α particle and a daughter nucleus can be determined as:

$$T_{1/2} = \frac{\ln 2}{\lambda} = \frac{\ln 2}{\nu P_\alpha}, \qquad (1)$$

where $\lambda$ is the decay constant and $\nu$ represents the assault frequency related to zero-point vibration energy $E_\nu$ as:

$$\nu = \frac{\omega}{2\pi} = \frac{2E_\nu}{h}. \qquad (2)$$

Empirical zero-point vibration energy $E_\nu$, in proportion to the released energy of emitted α particle $Q_\alpha$, can be calculated as [34]:

$$E_\nu = \begin{cases} 0.1045 Q_\alpha & \text{for even Z-even N parent nuclei,} \\ 0.0962 Q_\alpha & \text{for odd Z-even N parent nuclei,} \\ 0.0907 Q_\alpha & \text{for even Z-odd N parent nuclei,} \\ 0.0767 Q_\alpha & \text{for odd Z-odd N parent nuclei.} \end{cases} \qquad (3)$$

This simple formula for $E_\nu$ includes the pairing and shell effects of α decay.

The α-decay penetration probability $P_\alpha$ through the potential barrier can be calculated using the WKB semi-classical approximation as:

$$P_\alpha = \exp\left( \frac{-2}{\hbar} \int_{R_a}^{R_b} \sqrt{2\mu(V_T(r) - Q_\alpha)}\, dr \right), \qquad (4)$$

where $R_a$ and $R_b$ are the inner and outer turning points, respectively, determined as:

$$V_T(R_a) = Q_\alpha = V_T(R_b). \qquad (5)$$

### B. Description of potential formalism

Total interaction potential $V_T(r)$ between the emitted α particle and the daughter nucleus is taken to be the sum of the nuclear potential, Coulomb potential, and centrifugal potential as:

$$V_T(r) = V_N(r) + V_C(r) + V_l(r). \qquad (6)$$

Assuming homogeneous spherical charge distribution for the daughter nucleus, the Coulomb potential $V_C(r)$ between the α particle and a daughter nucleus using the point-like plus uniform model is:

$$V_C(r) = Z_\alpha Z_d e^2 \begin{cases} \dfrac{1}{r} & \text{for } r \geq R_c, \\ \dfrac{1}{2R_c}\left[3 - \left(\dfrac{r}{R_c}\right)^2\right] & \text{for } r \leq R_c, \end{cases} \qquad (7)$$

where $Z_\alpha = 2$ and $Z_d$ are the atomic numbers of the α particle and daughter nucleus, respectively, $r$ is the distance between the fragment centers, and $R_c$ denotes the touching radial separation between the α particle and the daughter nucleus. The rotational effects for the α-particle-daughter nucleus system can be calculated by the $l$-dependent centrifugal potential $V_l(r)$ as:

$$V_l(r) = \hbar^2 \frac{l(l+1)}{2\mu r^2}. \qquad (8)$$

The reduced mass of α-daughter system $\mu$ is:

$$\mu = m \frac{A_\alpha A_d}{A_\alpha + A_d}, \qquad (9)$$

where $m$ is the nucleon mass, $A_\alpha = 4$ and $A_d$ are the mass numbers of the α particle and the daughter nucleus, respectively, and $l$ is the orbital angular momentum carried away by the emitted α particle which is dictated by the spin-parity selection rule for α transition.

The nuclear part of the interacting potential is obtained by proximity formalism. The original version of the proximity potential for two spherical interacting nuclei was described by Blocki et al. [1] as:

$$V_N(r) = 4\pi \gamma b \bar{R} \Phi(\xi). \tag{10}$$

The first factors $(\gamma b \bar{R})$ refer to the geometry and shape of the participant nuclei and the remaining factor $(\Phi(\xi))$ is the universal function for separation distance $s$. Surface energy coefficient $\gamma$ is based on the Myers and Świątecki formula [9] and has the following form:

$$\gamma = \gamma_0 [1 - k_s A_s^2]. \tag{11}$$

Here, $A_s = \dfrac{N-Z}{N+Z}$ is the asymmetry parameter and refers to neutron/proton excess, where $N$ and $Z$ are the neutron and proton numbers of the parent nucleus, respectively, and $\gamma_0$ and $k_s$ are the surface energy constant and the surface asymmetry constant, respectively. The first set of constants was introduced by Myers and Świątecki [10] as $\gamma_0 = 1.01734$ MeV/fm$^2$ and $k_s = 1.79$ through the fitting of experimental binding energies. They then refined these constants to $\gamma_0 = 0.9517$ MeV/fm$^2$ and $k_s = 1.7826$ [9]. Blocki et al. used the more recent set of these constants in their formalism. This set is denoted herein as $\gamma$-MS 1967.

The width (diffuseness) of nuclear surface $b$ is considered close to unity ($b \approx 1$ fm). The mean curvature radius or reduced radius $\bar{R}$ in terms of matter radius $C_i$, the Süssmann's central radius, is:

$$\bar{R} = \frac{C_1 C_2}{C_1 + C_2}, \tag{12}$$

where:

$$C_i = R_i[1 - (\frac{b}{R_i})^2 + ...] \quad (i=1,2). \tag{13}$$

Effective sharp radius $R_i$ is defined as:

$$R_i = 1.28 A_i^{1/3} - 0.76 + 0.8 A_i^{-\frac{1}{3}} \text{ fm} \quad (i=1,2). \tag{14}$$

Here, index $i$ refers to the α particle and daughter nuclei. From this formula, the effective sharp radius of α particle $R_1$ is estimated to be 1.776 fm. The model succeeds provided that the diffuseness of the nuclei is much smaller than their radii.

The dimensionless universal function $\Phi(\xi = s/b)$, which only depends on separation distance $s$ between the half-density surfaces of the fragments, was obtained using the nuclear Thomas-Fermi model with Seyler-Blanchard phenomenological nucleon-nucleon interactions. The parameterization of the universal function is:

$$\Phi(\xi) = \begin{cases} \dfrac{-1}{2}(\xi - 2.54)^2 - 0.0852(\xi - 2.54)^3 & \text{for } \xi \leq 1.2511, \\ -3.437 \exp\left(-\dfrac{\xi}{0.75}\right) & \text{for } \xi \geq 1.2511, \end{cases} \tag{15}$$

where $\xi$ is the minimum separation distance in units of the surface width, and:

$$s = r - C_1 - C_2 \text{ fm}. \tag{16}$$

The universal function is independent of the shapes of two nuclei and geometry of the nuclear system. The proximity model was labeled as Proximity 1977 (Prox. 77).

In accordance with the different modifications on the adjustable parameters of Prox. 77, we have classified the proximity versions into eight major categories. These classifications are based on adjustments or changes in the surface energy coefficient, nuclei radius, and universal function. The sub-categories are presented in accordance with the smooth refinements on a special version. A total of 28 versions of proximity formalism are included in this study: (i) Prox. 77 family (included Prox. 77 [1] and its modified versions based on adjustment of the surface energy coefficient [9-17]), (ii) Proximity 1981 (Prox. 81) [2], (iii) Prox. 00 family (included Prox. 00 [3] and its modified forms Prox. 00DP [18], Prox.

2010 [19] and Dutt 2011 [20]), (iv) Bass family (included Bass 73 [4] and its modified forms Bass 77 [21] and Bass 80 [22]), (v) Winther family (included CW 76 [5] and its modified forms BW 91 [23] and AW 95 [24]), (vi) Ngô 80 [6], (vii) Denisov family (included Denisov [7] and its modified form Denisov DP [18]) and (viii) Guo 2013 [8]. Table I represents 13 different sets of the surface energy coefficient $\gamma(\gamma_0, k_s)$ which are applied on Prox. 77.

TABLE I. The different sets of the surface energy coefficient. $\gamma_0$ and $k_s$ are the surface energy constant and the surface asymmetry constant, respectively.

| $\gamma$ - set | $\gamma_0$ (MeV / fm$^2$) | $k_s$ | Ref. |
|---|---|---|---|
| set 1 ($\gamma$-MS 1967) | 0.9517 | 1.7826 | [9] |
| set 2 ($\gamma$-MS 1966) | 1.01734 | 1.79 | [10] |
| set 3 ($\gamma$-MN 1976) | 1.460734 | 4.0 | [11] |
| set 4 ($\gamma$-KNS 1979) | 1.2402 | 3.0 | [12] |
| set 5 ($\gamma$-MN-I 1981) | 1.1756 | 2.2 | [13] |
| set 6 ($\gamma$-MN-II 1981) | 1.27326 | 2.5 | [13] |
| set 7 ($\gamma$-MN-III 1981) | 1.2502 | 2.4 | [13] |
| set 8 ($\gamma$-RR 1984) | 0.9517 | 2.6 | [14] |
| set 9 ($\gamma$-MN 1988) | 1.2496 | 2.3 | [15] |
| set 10 ($\gamma$-MN 1995) | 1.25284 | 2.345 | [16] |
| set 11 ($\gamma$-PD-LDM 2003) | 1.08948 | 1.9830 | [17] |
| set 12 ($\gamma$-PD-NLD 2003) | 0.9180 | 0.7546 | [17] |
| set 13 ($\gamma$-PD-LSD 2003) | 0.911445 | 2.2938 | [17] |

## III. RESULTS AND DISCUSSION

### A. Systematic study of α-decay half-lives

A systematic comparative study was carried out on different versions of proximity formalism. The half-lives of α-decay ground-state-to-ground-state transitions of 344 nuclei with $52 \leq Z \leq 107$ were calculated within the framework of the WKB approximation. The selected α emitters were those with known values for their experimental α-decay half-lives for which the experimental value of energy released is available and considerable. A total of 136 even-even, 84 even-odd, 76 odd-even, and 48 odd-odd nuclei were considered. This nuclei set was the same as those used by Denisov and Khudenko [35], and Royer [36].

The total interaction potential between the α particle and the daughter nucleus was calculated as the sum of the nuclear proximity potential, Coulomb potential, and the centrifugal potential. Because of the spin-parity selection rule, the value of the orbital angular momentum for the ground-state-to-ground-state transitions was assumed equal to the minimum amount ($l = l_{min}$) that leads to the minimum centrifugal potential. The parent and daughter nuclei spin-parities and the $l_{min}$ for the emitted α particle were taken from Denisov and Khudenko [35], and Audi et al. [37], which were chosen from experimental data or data evaluation compilation analysis collections.

Twenty-eight versions of the proximity model were used to calculate the nuclear part of the total interaction potential between the α particle and the daughter nucleus. These versions result from modifications in different adjustable parameters in proximity formalism (Sec. II-B). The energies released by the α transition from the ground-state of the parent nucleus to the ground-state of daughter nucleus $Q_\alpha$ were taken from Wang et al. [38]. The turning points of the penetrability integral are the points at which the total interaction potential crosses the $Q_\alpha$ line and are determined using Eq. (5). The WKB barrier penetration probability is obtained by numerical solution of integral (4) between the inner and outer turning points.

The half-lives for α decay from the parent ground-state to the daughter ground-state were calculated using Eq. (1). The root-mean-square deviation (RMSD) was used to choose the best version of the proximity formalism that leads to acceptable half-lives for α decay as:

$$\text{RMSD} = \sqrt{\frac{1}{n}\sum_{i=1}^{n}\left[log_{10}\left(\frac{T_{1/2\ i}^{cal}}{T_{1/2\ i}^{exp}}\right)\right]^2}, \quad (17)$$

where $T_{1/2}^{cal}$ and $T_{1/2}^{\exp}$ are the calculated and experimental α-decay half-lives, respectively, and parameter $n$ denotes the number of nuclei included in the summation that have definite theoretical half-lives. The experimental data for the α-decay half-lives are the same as those used by Denisov and Khudenko [35]. The RMSDs of the decimal logarithmic half-lives versus the various versions of the proximity potential are presented in Table II. The numbers in parentheses refer to the number of nuclei under consideration.

The total potential barrier for some models is where line $Q_\alpha$ is located under the touching configuration and there is no inner turning point; hence, the penetration probability and the α half-life cannot be determined. For some models, this means that calculations are limited to fewer isotopes, especially in the Prox. 77-set 1, Prox. 77-set 8, Prox. 77-set 12, Prox. 77-set 13, and Prox. 00 models. These models were discarded because their number of nuclei under calculation was ≤282. The calculated half-lives for most other versions cover all 344 nuclei. Most models give deviations of <1.1. The best results were obtained by the Prox. 77-set 4, Prox. 77-set 5, and Dutt 2011 models with RMSDs of 0.99, 0.96, and 0.96, respectively. The RMSDs derived by potentials affected by Prox. 00DP, Prox. 2010, Denisov, and Denisov DP were high; therefore, these models are not appropriate in their original forms for α-decay calculations.

The RMSDs for all proximity models were recalculated with respect to the even-even, even-odd, odd-even, and odd-odd α emitters. The results are presented in Table II along with the total data. Table II indicates that the RMSDs evaluated using most proximity versions decreased strongly for even-even nuclei in comparison with odd-even, even-odd, and odd-odd nuclei. For even-even nuclei, the α half-lives predicted by the prox. 77-set 3, prox. 77-set 4, prox. 77-set 5, prox. 77-set 6, prox. 77-set 7, prox. 77-set 9, prox. 77-set 10, Dutt 2011, Bass 73, Bass 80, and BW 91 models were very close to the experimental results. The RMSDs related to these cases were 0.54 to 0.60. Although, most versions were not very efficient for the even-odd, odd-even, and odd-odd nuclei, the results were within acceptable range. It appears that the theoretical proximity model worked much better for even-even isotopes than for the others.

It should be pointed out that the results were very sensitive to the value of released energy. Möller et al. [39] stated that an uncertainty of 1 MeV in the $Q_\alpha$ value corresponds to an uncertainty of the α-decay half-life of $10^3$ to $10^5$ times in the heavy element region. The choice of different experimental masses and $Q_\alpha$ values from different references or application of different formulas to calculate $Q_\alpha$ will affect the results. Moreover, the values for the experimental α-decay half-lives differ in different studies according to the methods applied. It is noteworthy that the experimental values update continually and the number of known α emitters is expanding with advancements in instruments and methods.

**B. Comparison with other investigations**

To measure the acceptability of the proximity models for α-decay studies, the results were compared with those from other approaches. Denisov and Khudenko [35] evaluated the ground-state-to-ground-state α-decay half-lives for a nuclei set as was done in the present study. They calculated the α-decay half-lives in the framework of the unified model for α decay and α capture (UMADAC) by considering the spin-parity effect and deformation of the parent and daughter nuclei. The total potential was a function of $(r,\ \theta,\ l,\ Q_\alpha)$. They tabulated their results with the results from various empirical relationships [39,40,41] in which, based on the fitting parameters and special analytical parameters, $log_{10}(T_{1/2})$ was expressed as a simple function of α-particle energy, charge, and mass of the parent nuclei. The relationships were based on a pure Coulomb potential and neglected the nuclear potential, deformation, and spin-parity effects. The α half-lives taken from Denisov 's UMADAC were in good agreement with the experimental values.

Royer [36] applied the generalized liquid-drop model (GLDM), including the proximity effects on the same data set. He proposed analytical formulas for $log_{10}(T_{1/2})$ that either depend or do not depend on angular momentum (i.e., $log_{10}(T_{1/2}) = \text{f}(Q_\alpha,\ A,\ Z)$ or $log_{10}(T_{1/2}) = \text{f}(Q_\alpha,\ l,\ A,\ Z)$). Royer stated that entering $l$-dependence in $log_{10}(T_{1/2})$ formulas improved the results. The RMSDs for α half-lives, except for nuclei such as $^{113}_{53}\text{I}$, $^{149}_{64}\text{Gd}$, $^{206}_{85}\text{At}$, $^{218}_{91}\text{Pa}$, and $^{235}_{95}\text{Am}$ were relatively small. He concluded that, in the case of these exceptions, the related experimental half-lives can be disputed.

Zhang et al. [31] applied the proximity formalism derived from Guo et al. [8] using modified parameters to study α-decay half-lives of 145 heavy nuclei with 52 ≤ Z ≤ 92. Their universal function was formulated by a systematic study using the double folding model with density-dependent nucleon-nucleon interaction (CDM3Y6). This model is referred to as Zhang 2013. The present study extended their calculations to 344 nuclei.

Table III lists the RMSDs of the decimal logarithm of the α-decay half-lives calculated using three versions of the proximity model (Prox. 77-set 4, Prox.77-set 5, and Dutt 2011) for comparison with UMADAC, other empirical approaches, and the Zhang proximity formula. The table reveals that, although effects such as deformations are not included in proximity versions, the results are comparable with the results of Denisov and Royer and are better than other approaches. It is notable that the values of $Q_\alpha$ differed at times in these investigations.

## IV. CONCLUSION

The α-decay half-lives of nuclei with atomic numbers in the range of 52 ≤ Z ≤ 107 were estimated from the WKB penetration probability through potential barriers. A systematic comparative study was performed on 28 versions of the proximity potential used to calculate the nuclear part of the potential. The theoretical results were compared with the experimental data using the RMSD. The RMSD of the decimal logarithm of the half-lives had the lowest value for the proximity model with versions Prox. 77-set 4, Prox. 77-set 5, and Dutt 2011. For the same data set, the RMSDs were recalculated for the even-even, even-odd, odd-even, and odd-odd nuclei. The results revealed that, for most models, the RMSDs decreased strongly for the even-even parent nuclei. It appears that the proximity model is more applicable to even-even nuclei.

The results evaluated using proximity models were compared with those obtained using the UMADAC derived from Denisov, other empirical approaches, and the Zhang proximity formula. The results were comparable to or better than the other approaches. It appears that of 28 proximity models studied, Dutt 2011 was able to perfectly predict both α-decay half-lives and fusion cross sections.

It should be noted that the challenge to discovery of the best proximity model has not been fully met. It is possible to improve upon the best versions of the proximity potential for α decay by incorporating new modifications such as deformation and orientation effects to the nuclear and Coulomb potentials, defining an appropriate potential for the overlapping region, considering fine-structure, and incorporating the preformation factor into the half-life formula.

TABLE II. RMSDs of the decimal logarithm of α-decay half-lives for different versions of the proximity potential. The data set consists of 344 total, 136 even-even, 84 even-odd, 76 odd-even, and 48 odd-odd parent nuclei. The numbers in parentheses are the number of nuclei under consideration.

| Proximity model | Total | even-even | even-odd | odd-even | odd-odd |
|---|---|---|---|---|---|
| Prox. 77-set 1 | 0.7515 (262) | 0.8458 (100) | 0.6780 (68) | 0.6525 (53) | 0.7427 (41) |
| Prox. 77-set 2 | 0.8193 (324) | 0.7396 (128) | 0.9014 (77) | 0.8755 (72) | 0.7951 (47) |
| Prox. 77-set 3 | 1.1463 (344) | 0.5900 (136) | 1.4891 (84) | 1.3595 (76) | 1.2742 (48) |
| Prox. 77-set 4 | 0.9861 (344) | 0.5605 (136) | 1.2797 (84) | 1.1522 (76) | 1.0543 (48) |
| Prox. 77-set 5 | 0.9625 (344) | 0.5684 (136) | 1.2457 (84) | 1.1185 (76) | 1.0136 (48) |
| Prox. 77-set 6 | 1.0432 (344) | 0.5452 (136) | 1.3640 (84) | 1.2341 (76) | 1.1353 (48) |
| Prox. 77-set 7 | 1.0252 (344) | 0.5451 (136) | 1.3397 (84) | 1.2100 (76) | 1.1099 (48) |
| Prox. 77-set 8 | 0.7622 (229) | 0.8464 (89) | 0.7119 (58) | 0.6547 (48) | 0.7540 (34) |
| Prox. 77-set 9 | 1.0295 (344) | 0.5440 (136) | 1.3461 (84) | 1.2162 (76) | 1.1156 (48) |
| Prox. 77-set 10 | 1.0305 (344) | 0.5443 (136) | 1.3473 (84) | 1.2174 (76) | 1.1171 (48) |
| Prox. 77-set 11 | 0.8955 (341) | 0.6550 (136) | 1.1060 (82) | 1.0103 (75) | 0.8924 (48) |
| Prox. 77-set 12 | 0.7797 (266) | 0.8721 (102) | 0.7283 (70) | 0.6693 (53) | 0.7533 (41) |
| Prox. 77-set 13 | 0.8634 (183) | 0.9320 (70) | 0.8387 (47) | 0.8016 (40) | 0.8048 (26) |
| Prox. 81 | 0.9293 (341) | 1.0058 (136) | 0.9421 (82) | 0.8845 (75) | 0.7281 (48) |
| Prox. 00 | 0.7349 (282) | 0.5442 (109) | 0.8184 (73) | 0.7780 (58) | 0.9260 (42) |
| Prox. 00DP | 2.2758 (344) | 1.6712 (136) | 2.6742 (84) | 2.5660 (76) | 2.5032 (48) |
| Prox. 2010 | 2.3885 (344) | 1.7907 (136) | 2.7857 (84) | 2.6767 (76) | 2.6220 (48) |
| Dutt 2011 | 0.9639 (344) | 0.5517 (136) | 1.2528 (84) | 1.1238 (76) | 1.0246 (48) |
| Bass 73 | 1.1744 (344) | 0.5978 (136) | 1.5189 (84) | 1.3871 (76) | 1.3371 (48) |
| Bass 77 | 2.0314 (344) | 2.3842 (136) | 1.7457 (84) | 1.8312 (76) | 1.6811 (48) |
| Bass 80 | 1.0765 (344) | 0.5425 (136) | 1.4095 (84) | 1.2790 (76) | 1.1849 (48) |
| CW 76 | 1.6206 (344) | 0.9771 (136) | 2.0213 (84) | 1.9045 (76) | 1.7956 (48) |
| BW 91 | 1.0464 (344) | 0.5351 (136) | 1.3704 (84) | 1.2405 (76) | 1.1462 (48) |
| AW 95 | 1.4027 (344) | 0.7684 (136) | 1.7862 (84) | 1.6610 (76) | 1.5736 (48) |
| Ngô 80 | 1.0750 (342) | 1.2003 (136) | 1.0505 (83) | 0.9977 (75) | 0.8296 (48) |
| Denisov | 2.8956 (344) | 2.2984 (136) | 3.2986 (84) | 3.2230 (76) | 3.1040 (48) |
| Denisov DP | 3.9536 (344) | 3.3830 (136) | 4.3466 (84) | 4.2925 (76) | 4.1665 (48) |
| Guo 2013 | 1.9040 (344) | 1.2693 (136) | 2.3040 (84) | 2.1957 (76) | 2.1196 (48) |

TABLE III. Comparison of RMSDs of the decimal logarithm of α-decay half-lives for a full set of nuclei derived using different approaches. All investigations employ 344 total, 136 even-even, 84 even-odd, 76 odd-even, and 48 odd-odd α emitters.

| model | Total | even-even | even-odd | odd-even | odd-odd | Ref. |
|---|---|---|---|---|---|---|
| Prox. 77-set 4 | 0.9861 | 0.5605 | 1.2797 | 1.1522 | 1.0543 | present |
| Prox. 77-set 5 | 0.9625 | 0.5684 | 1.2457 | 1.1185 | 1.0136 | present |
| Dutt 2011 | 0.9639 | 0.5517 | 1.2528 | 1.1238 | 1.0246 | present |
| Denisov: UMADAC | 0.6199 | 0.2980 | 0.7805 | 0.7613 | 0.7405 | [35] |
| Dasgupta-Schubert | 1.0245 | 0.5205 | 1.1661 | 1.3453 | 1.2617 | [40] |
| Medeiros | 1.1344 | 0.3652 | 1.5510 | 1.3635 | 1.3390 | [41] |
| Möller | 1.3926 | 1.3067 | 1.4389 | 1.5728 | 1.2828 | [39] |
| Royer: $l$-independent | 0.7452 | 0.3280 | 0.9559 | 0.8891 | 0.9080 | [36] |
| Royer: $l$-dependent | 0.5296 | 0.3280 | 0.5552 | 0.6661 | 0.6807 | [36] |
| Zhang 2013 | 1.3764 | 0.7373 | 1.7578 | 1.6348 | 1.5483 | [31] |